\documentclass[preprint,prc,aps,showkeys]{revtex4-2}

\usepackage{graphicx}
\usepackage{xcolor}

\usepackage{amsmath} 
\usepackage{amssymb} 
\usepackage{subfig}   
\usepackage{multirow}
\usepackage{inputenc}
\usepackage{soul} 
\usepackage{natbib}
\usepackage{graphics}
\usepackage{array}
\usepackage{multirow}

\usepackage{float}
\usepackage{subfig}
\usepackage{graphicx}
\usepackage{xcolor}

\usepackage{booktabs}

\begin{document}

    \title{The $D_sDK^*$  vertex in QCD Sum Rules: form factors and coupling constant}

    \author{A. Cerqueira Jr$^a$} \email{angelocunha@uezo.rj.gov.br}
    \author{B. Os\'orio Rodrigues$^b$}\email{bruno.osorio.rodrigues@uerj.br}
    \author{M. E. Bracco$^c$}\email{mirian.bracco@fat.uerj.br}
    \author{C. M. Zanetti$^c$}\email{carina.zanetti@fat.uerj.br}

      \affiliation{$^a$ Escola de Engenharias, Centro Universit\'ario Estadual da Zona Oeste, Av. Manuel Caldeira de Alvarenga, 1203, 23070-200, Rio de Janeiro, RJ, Brazil.}
    \affiliation{$^b$ Instituto de Aplica\c{c}\~ao Fernando Rodrigues da Silveira, Universidade do Estado do Rio de
        Janeiro, Rua Santa Alexandrina 288, 20261-232, Rio de Janeiro, RJ, Brazil. }
    \affiliation{$^c$ Faculdade de Tecnologia, Universidade do Estado do Rio de Janeiro,
        Rod. Presidente Dutra Km 298, P\'olo Industrial, 27537-000, Resende, RJ, Brazil.}

    \begin{abstract}
        In this work we study the meson vertex $D_sDK^*$ using the QCD sum rules. We compute the three point correlation functions for the three cases of different off-shell mesons. The form factors for each case are fitted to the numerical calculation of the correlation functions and the coupling constant of the vertex is obtained by comparing the form factors at their respective off-shell meson pole.
        The result  obtained for the  coupling constant  is $g_{D_sDK^*} = 2.29^{+0.65}_{-0.41}$.

    \end{abstract}

    \keywords{QCD Sum Rules; Strong coupling constant; Form factor; Three   meson vertex.}

\maketitle

\newpage
\section{Introduction}

The study of hadrons has widely improved by the efforts of many experimental collaborations such as Belle (KEK, Japan), BaBar (SLAC, USA), BESIII (BEPCII, China), CDF  and LHCb (LHC, Switzerland), that have been collecting a large amount of experimental data  in the last decades. These data  have become a remarkable testing ground for  different approaches that deal with the non-perturbative nature of the strong interactions between hadrons in decay processes. The calculation of the cross sections of the decay processes can be used to elucidate a little more about the complex world of the fundamental particles interactions. In order to study these hadronic decay processes, it is necessary to address the large theoretical challenge of the non-perturbative nature of Quantum Chromodynamics (QCD). 
The QCD, the fundamental theory of the strong interactions between quarks and gluons, is non-perturbative at low energies since its coupling constant $\alpha_s(q^2)$ increases as $|q^2|$ decreases. Therefore, it is necessary to use some non-perturbative approaches like some effective field theories. One of the more usual approach used is based in the effective $SU(4)$ lagrangian,  \cite{Bracco:2011pg,Lin:2000,Oh:2001}, where the quarks $u$, $d$, $s$ and $c$ are considered to have the same mass.
With this effective theory, it is possible to perform the theoretical cross section, where two fundamental ingredients are needed, one is the form factor and the other is the coupling constant of the vertex. The form factor is responsible to carry the unknown non-perturbative dynamics of the strong interactions. It is represented by a function of the squared momentum,  $g(q^2)$, and is not always directly observed experimentally. For this reason, the shape of the form factor used has the same shape as low energy nuclear physics, where only the parameters need to be adjusted. The situation is similar for the coupling constants, there is not, in general, experimental data that permit to infer their  values. One exception is the coupling constant of the  $D^* D \pi$ decay, where through the width of $D^*$,
the decay  coupling constant was calculated \cite{Navarra:2001ju}. But in other cases, the coupling constant would be obtain using the relations of the $SU(4)$ \cite{Bracco:2011pg}.

Our group had been improving a method to calculate the form factors and coupling constants, in some charmed and botton decays,  using a non-perturbative approach called Quantum Chromodynamics Sum Rules (QCDSR).  
The QCDSR technique was developed by the authors Shifman, Vainshtein and Zakharof \cite{Shifman:1978by,Shifman:1979}. It is a powerful and analytical tool that allow us to compute the physical properties of hadrons, such as the masses, decay constants, form factors and coupling constants.  
We have been studying, for a long time, a technique to obtain form factors and coupling constant  and improve the results with less uncertainties. Some of our three mesons strange-charmed vertices that were calculated are: $J/\psi D_s D_s$ \cite{Rodrigues:2013bta}, $J/\psi D_{s}^{*} D_{s}^{*}$ \cite{Rodrigues:2020anj}, $\eta_c D^* D$ and $\eta_c D_s^* D_s$ \cite{Rodrigues:2017qsm}, $ D_s D_ s \phi $ \cite{OsorioRodrigues:2017aqv}, $D^* D_s K$ and $ D^*_s D K $ \cite{Bracco:2006xf}, the $B_{s}B^{*}K$ \cite{Cerqueira:2012zsa} , the $B_s B^* K $ and $B_s B K^*$ \cite{Cerqueira:2015vva},  and  many other that can be seen in Ref.~\cite{Bracco:2011pg}. 

Our method consists of calculating, for the three meson decay, three different form factors that are obtained when each meson is used as a projectile of the vertex (the target). In this way, we obtain information about how each particle "reads"  the interaction vertex. But each form factor, $g_i(q^2)$, may give the same value for the coupling constant, which could be obtained when $q^2\to m^2_{\rm{off}}$, where $m_{\rm{off}}$ is the mass of the off-shell meson. This is possible when we work in the QCDSR respecting the stability of the sum rule. 
The details of our approach and the earlier results are reviewed in Ref.~\cite{Bracco:2011pg}. In that review, it is noted that the shape of the form factors in several charmed decay processes present some regularity, related to the shape of the function that represent the form factor. 

\begin{figure}[h!]
    \centering
  {\includegraphics[width = 0.4\textwidth]{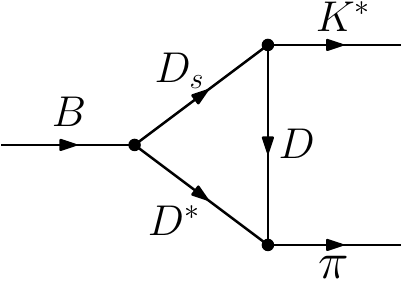}}
    \caption{FSI decay process $B \to DD_s \to K^*\pi$.}
    \label{fig-intro}
\end{figure}

Following up our studies of  vertices with open charm mesons, our objective in the present work is to apply our method to study the form factors and the coupling constant of the vertex $D_sDK^*$. In a previous work, the vertices $D^*D_s K$ and $DD_s^*K$ were calculated \cite{Bracco:2006xf}. These vertices are important in studies of the dissociation cross section of $J/\psi$ by kaons, and the suppression of the charmonium production is an important signature of the quark-gluon plasma (QGP) in heavy ion collisions. Another important type of processes that these vertices plays a role are decays with final state interactions (FSI), in which a meson decays into an intermediate state, and then the mesons of the intermediate state interact via particle exchange, before decaying into the final state. One example of such process is the $B^\pm\to DD_s\to K^*\pi^\pm$, shown in Fig.~(\ref{fig-intro}). This decay is important because it can be used to detect direct CP violation and to determine the CKM parameters mixing. Note that the vertex $BDD_s$ represents an effective electroweak interaction, so it can be evaluated by the Hamiltonian of the electroweak theory \cite{Isola:2003fh}.

Considering that the vertex $D_s D K^*$ has three distinct mesons,  each one of them can act as the projectile. Therefore, it is necessary to take this into account and compute the three-point correlation function for all the possible cases, obtaining three different form factors and coupling constants. Given that the coupling constant must be the same regardless of which meson is off-shell, the mean value of the three calculated couplings will result in a single value for the coupling constant of the vertex. This procedure gives a more reliable result, since there is no reason to choose only one of three possible form factors to determine the coupling constant. Besides that, an important source of uncertainty in the numerical calculation originates from the fact that the QCDSR results are reliable on a small window of momentum, with $Q^2>0$, and the coupling constant is obtained from the value of the form factor at the off-shell meson pole, $g=g(Q^2=-m_{\rm{off}}^2)$ (i.e. at $Q^2<0$), meaning that the results have to be fitted in this small window and then be extrapolated to a range of momentum outside the window which the sum rules calculations are actually performed. Therefore, by calculating all the possible off-shell cases, the uncertainties regarding this extrapolation are also minimized by the procedure, as the three coupling constants calculated must converge to a single value (considering their error bars).

This paper is organized as follows: in section II, we show the calculation of the three-point correlation function of the vertex $D_sDK^*$; in section III, we present the final equations of the sum rule for each off-shell case; the numerical analysis and results are discussed in section IV; our final remarks and conclusion are on section V, where we also compare our results with previous calculations found in literature.

\section{The three point correlation function}

The main element of the QCDSR technique is the correlation function, such as the two-point correlation function that can be used to obtain the mass of the hadron, and the three-point function used to obtain the form factors and coupling constants. The quark-hadron duality principle states that hadrons can be described equally by the quarks and gluons degrees of freedom and also by the hadronic degrees of freedom. Thus the correlation function can be computed in two different ways: on the OPE side (or QCD side), the quark and gluons fields are used; and on the phenomenological side the hadronic ones are used. When comparing both sides of the sum rule, it is possible to obtain an equation for the physical parameters that are introduced on the phenomenological side.

The three-point correlation function for the vertex $D_sDK^*$,  with the off-shell meson $M=D_s,D,K^*$, is written as: 
\begin{equation}
\Gamma_\mu^{(M)}=\int d^4x\,d^4y\,e^{ip^\prime\cdot x}\,e^{-iq\cdot y}\langle 0\vert T{j_1(x)j_2^\dagger(y)j_3^\dagger(0)}\vert 0\rangle,
\label{correlator}
\end{equation}
where the relation between the mesons four momentum is
$q=p^\prime-p$, and the functions $j_i$ are meson interpolating currents that represent the mesons in terms of quarks degrees of freedom, when it is computed by the QCD side.

\begin{figure}[h!]
    \subfloat[$D$ off-shell]{\includegraphics[width = 0.3\textwidth]{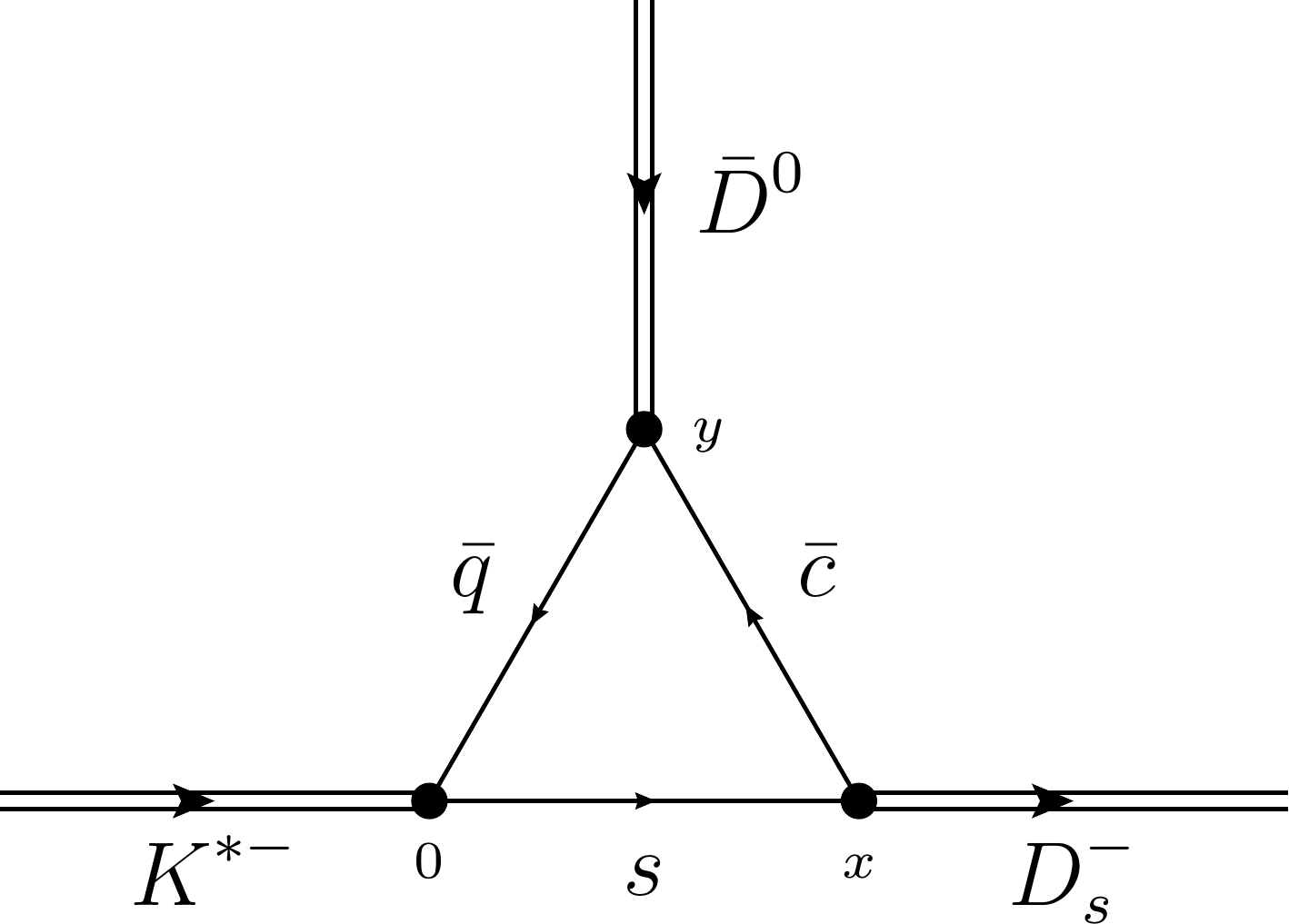}}\qquad
    \subfloat[$D_s$ off-shell]{\includegraphics[width = 0.3\textwidth]{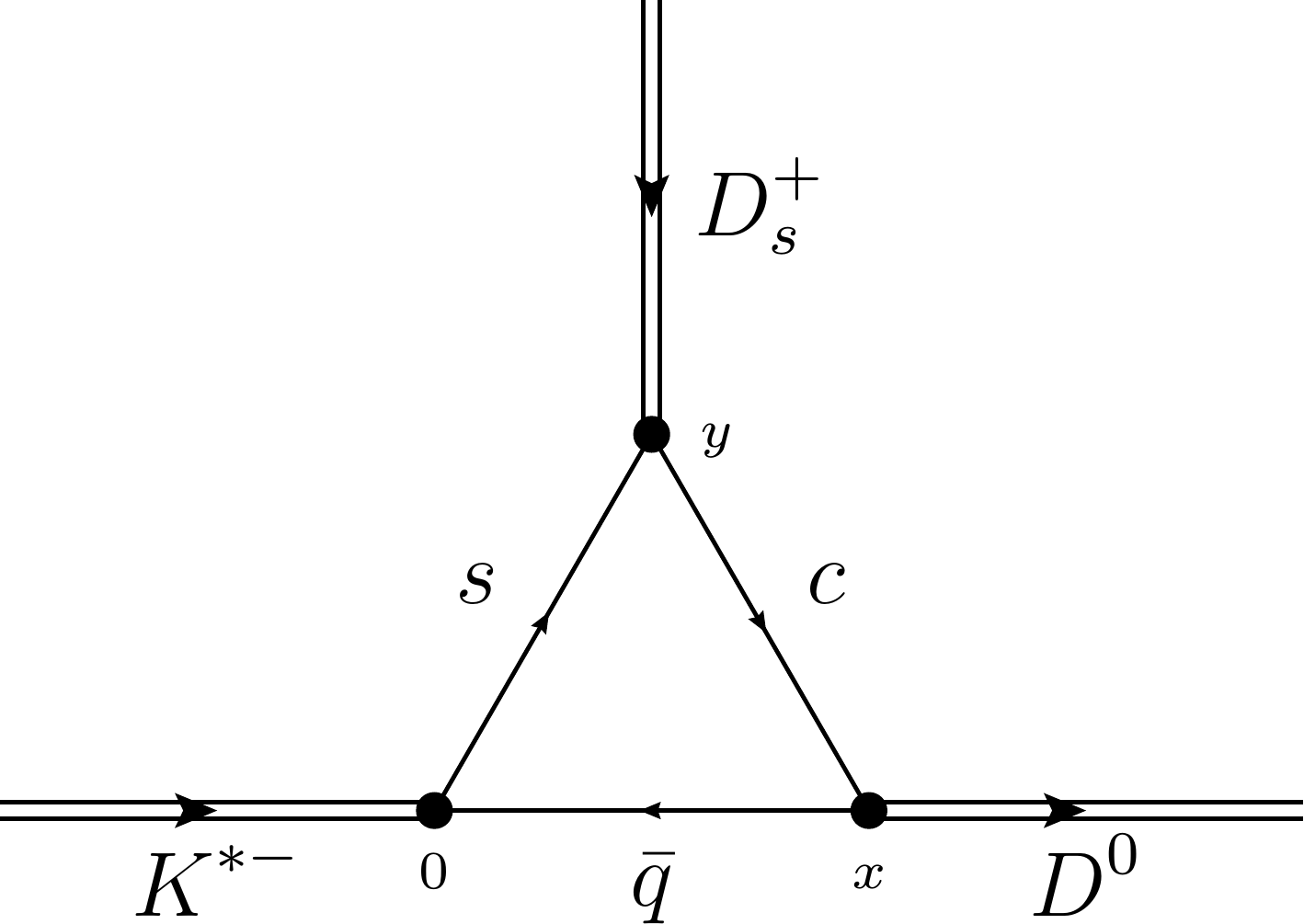}}\qquad
    \subfloat[$K^*$ off-shell]{\includegraphics[width = 0.3\textwidth]{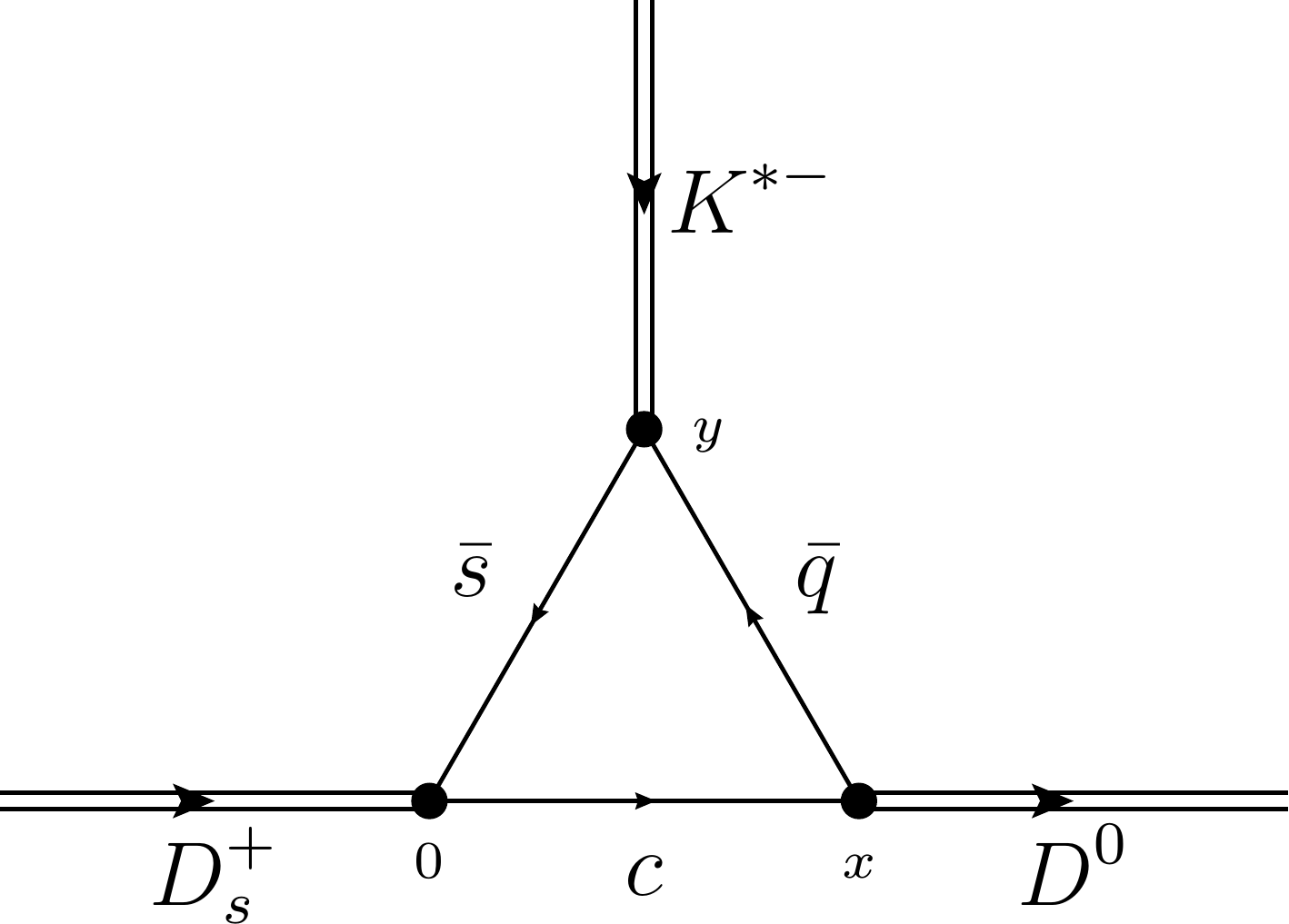}}
    \caption{The   $D_sDK^*$ vertex  with  off-shell mesons.}
    \label{diagrams}
\end{figure}

The order of the currents on the correlator of
Eq.~(\ref{correlator}) depends on the  mesons configuration in
the vertex \cite{Bracco:2006db,Cerqueira:2011za}. We will study
the three possible cases of off-shell mesons for this vertex --
$D$, $D_s$ and $K^*$ off-shell -- as shown in
Fig.~(\ref{diagrams}).

For this vertex, we need the following currents that contain the quantum properties  of the mesons written in terms of quark fields:
\begin{eqnarray}
j_\mu^{K^*}&=&\bar{u}\gamma_\mu s,\nonumber\\
j^D_5&=&i\bar{c}\gamma_5 q,\nonumber\\
j^{D_s}_5&=&i\bar{c}\gamma_5 s, \label{3currents}
\end{eqnarray}
where $q$ is the up or down quark, $c$ is the charm and $s$ is the strange quark.

On the OPE side of the sum rule, the correlation function is calculated by the input of the interpolating currents from Eq.~(\ref{3currents}) into the Eq.~(\ref{correlator}), and then  a Wilson operator product expansion (OPE) is carried out. Here, the time ordered product of the local currents is expanded in terms of local operators and incorporates the non-perturbative effects corrections in the sum rule. These operators are ordered in terms of increasing dimension: the first one is the unitary operator for the perturbative contribution; the next ones are non-perturbative contributions composed by quarks and gluons fields and their  QCD vacuum expectation values, known as the condensates of quarks and gluons.
 
Using the Cutkosky's rule, the correlation function can be written as  a double dispersion relation, for the different Dirac structures present in the function:
\begin{equation}
\Gamma_\mu^{\rm OPE}(p^2,p^{\prime2},q^2)=-\frac{1}{4\pi^2}\int_{0}^{\infty}\int_{0}^{\infty} ds\,du\frac{\rho_\mu^{\rm OPE}(s,u,q^2)}{(s-p^2)(u-p^{\prime 2})},
\end{equation}
where the function $\rho_\mu^{\rm{OPE}}$ is the spectral density that is
related to the double discontinuity of the amplitude.  The
spectral densities are computed at first order of the perturbative
expansion in $\alpha_s$, the strong coupling constant. The OPE in this work includes terms
up to dimension five of the condensates, leading to the  following OPE side of the sum rule:
\begin{equation}
\Gamma_\mu^{\rm OPE}=\Gamma_\mu^{\rm pert}+\Gamma_\mu^{\langle\bar{q}q\rangle}+\Gamma_\mu^{m_s\langle\bar{q}q\rangle}+\Gamma_\mu^{\langle g_s^2G^2\rangle}+\Gamma_\mu^{\langle\bar{q}Gq\rangle}+\Gamma_\mu^{m_s\langle\bar{q}Gq\rangle},
\end{equation}
where $q$ is a quark up, down or strange, $\Gamma_\mu^{\rm  pert}$ is the perturbative term of the OPE, the following terms are the non-perturbative contributions from quarks $\langle\bar{q}q\rangle$, gluons $\langle g_s^2G^2\rangle$ and mixed condensates $\langle\bar{q}Gq\rangle$ and $m_s$ mass corrections.

On the phenomenological side of the sum rule, the correlation
function of Eq.~(\ref{correlator}) is interpreted in terms of
meson currents that creates and annihilates the hadronic states of
the vertex. This is done by inserting the intermediate hadronic
states in Eq.~(\ref{correlator}). The  matrix elements thus
obtained are written in terms of the hadronic physical parameters -- the
masses and decay constants:
\begin{eqnarray}
&&\langle 0\vert j_{D}\vert D(p_a)\rangle=\frac{m_{D}^2f_{D}}{m_c+m_u}\,,\nonumber\\
&&\langle 0\vert j_{D_s}\vert D_s(p_b)\rangle=\frac{m_{D_{s}}^2f_{D_s}}{m_c+m_s}\,,\nonumber\\
&&\langle 0\vert j_\mu^{K^*}\vert K^*(p_c,\epsilon)\rangle=m_{K^*}f_{K^*}\epsilon_\mu(p_c)\,.\label{massdecay}
\end{eqnarray}
The respective mesons four momenta $p_a$, $p_b$ and $p_c$ can be
$p$, $p^\prime$ or $q$, depending on which meson is off-shell,
according to the Fig.~(\ref{diagrams}). The vertex amplitude
matrix is derived from the hadronic interaction lagrangian, which
defines the coupling constant of the vertex ${D_sDK^*}$
\cite{Azevedo:2003qh}:
\begin{equation}
\mathcal{L}=ig_{D_sDK^*}\left[K^{*\alpha}\left(D\partial_\alpha\bar{D_s}-\bar{D_s}\partial_\alpha D\right)+\bar{K}^{*\alpha}\left(D_s\partial_\alpha\bar{D}-\bar{D}\partial_\alpha D_s\right)\right].\label{lagrangian}
\end{equation}
The amplitude matrices that appear in the correlation function for each off-shell meson $M$ are extracted from Eq.~(\ref{lagrangian}), and are given in terms of the form factors $g_{D_sDK^*}^{(M)}(q^2)$.  The resulting amplitude matrices for the $D$, $D_s$ and $K^*$ off-shell meson cases are then written respectively as:
\begin{eqnarray}
&&\langle D(q)\vert K^*(p)D_s(p^\prime)\rangle=ig_{D_sDK^*}^{(D)}(q^2)(q_\alpha+p_\alpha^\prime)\epsilon^\alpha(p)\,,\nonumber\\
&&\langle D_s(q)\vert K^*(p)D(p^\prime)\rangle=-ig_{D_sDK^*}^{(D_s)}(q^2)(q_\alpha+p_\alpha^\prime)\epsilon^\alpha(p)\,,\nonumber\\
&&\langle K^*(q)\vert D_s(p)D(p^\prime)\rangle=-ig_{D_sDK^*}^{(K^*)}(q^2)(p_\alpha+p_\alpha^\prime)\epsilon^\alpha(q)\,,\label{matrices}
\end{eqnarray}
where the $g^{(M)}(q^2)$'s functions are the respective form factors of the vertices.


\section{Sum rules}

The vertices sum rules are lastly obtained following the quark-hadron duality principle, by which the two sides of the sum rule  can be matched. The possible Dirac structures for the correlation functions of the $D_sDK^*$ vertex are $p_\mu$ and $p_\mu^\prime$.  Matching the respective OPE and phenomenological sides of the sum rules, we obtain a sum rule for every Dirac structure of each one of the off-shell meson cases.

In order to improve the  matching of the sides of the sum rule, a double Borel
transformation is performed on both sides of the sum rules. This
operation introduces the Borel mass parameters $M$ and $M^\prime$.
The reason for this step in the calculation is that the Borel
transformation enhances the convergence on the OPE side and
suppresses higher order terms. After equating the OPE and
phenomenological sides, the double Borel transformation is
performed, with $P^2\to M^2$ and $P^{\prime 2} \to M^{\prime 2}$,
\begin{equation}
\mathcal{BB}\left[\Gamma^{\rm OPE}_{i}(P^2,P^{\prime 2},Q^2)\right]=\mathcal{BB}\left[\Gamma^{\rm phen}_i(P^2,P^{\prime 2},Q^2)\right],
\end{equation}
where the squared mesons four momenta is in the Euclidean
space, $p^2=-P^2$, $p^{\prime 2}=-P^{\prime 2}$  and $q^2=-Q^2$.
The functions $\Gamma_i$ represent the correlation functions for
each Dirac structure ($i = p_\mu, p_\mu^\prime$).

At this point, the last step to assure the matching of both sides
of the sum rules, is to introduce the threshold parameters
$s_{sup}$ and $u_{sup}$ in the limits of integration on the OPE
side, separating the pole from the continuum contributions. When
this is done, only the pole contribution remains on both sides of
the sum rule: the OPE continuum contribution cancels out with the
h.r. terms (corresponding to the excited states above the pole)
on the phenomenological side. The threshold parameters are defined
as $s_{sup}=(m_i+\Delta_i)^2$ and $u_{sup}=(m_o+\Delta_o)^2$, with
the masses $m_i$ and $m_o$ being the masses of the incoming and
outgoing mesons of the vertex according to Fig.~(\ref{diagrams}).
The $\Delta_i$ and $\Delta_o$ parameters represent the gap between
the pole resonances ($M=D,D_s,K^*$) and their excited states. The values of the $\Delta$'s are
determined by the best fit in the numerical analysis, and it is
usually found to be around $0.5$ GeV.

The final expressions for the sum rules can be written by isolating the form factors in their respective equations and rewriting the phenomenological side as $\mathcal{B}\mathcal{B}[{\Gamma}^{\rm phen}_{(i)}(Q^2)]$ $=g(Q^2)\cdot\mathcal{B}\mathcal{B}[\widetilde{\Gamma}^{\rm phen}_{(i)}(Q^2)]$. The resulting general expression is given bellow:
\begin{align}
g_{D_sDK^*}^{(M)}(Q^2)  =  \frac{-  \frac{1}{4\pi^2}
    \int^{s_{sup}}_{s_{inf}}\int^{u_{sup}}_{u_{inf}}
    DD\left[\Gamma^{\rm pert}_{i}\right]       e^{-\frac{(s+u)}{M^2}}dsdu      +
    \mathcal{B}    \mathcal{B}\left     [\Gamma^{\rm non-pert}_{i}    \right]
}{\mathcal{B}    \mathcal{B}\left   [\widetilde{\Gamma}_{i}^{\rm phen}(Q^2)
    \right ] }\,.\label{formfactorfunction}
\end{align}
for the $M=D,D_s$ and $K^*$ off-shell cases.

The non-perturbative terms of the OPE ($\Gamma^{\rm non-pert}$) correspond to the condensates  contributions for each off-shell case: quark, gluon and mixed for the $D$ and
 $D_s$ off-shell cases, and only gluon condensates for the $K^*$ off-shell case. The expressions for the OPE terms and the limits of integration can be found in the appendix A.

 We have also applied the following ansatzes for the Borel masses parameters, for the $D,D_s$ and $K^*$ off-shell, respectively:
 \begin{equation}
 M'^2=M^2 \frac{m^2_{D_s} - m_c^2}{m_{K^*}^2}
 \end{equation}
 \begin{equation}
 M'^2=M^2 \frac{m^2_D - m_c^2}{m_{K^*}^2}
 \end{equation}
 \begin{equation}
 M'^2=M^2 \frac{m_D^2}{m_{D_s}^2}
 \end{equation}

\section{Results}

\begin{table}[h]
    \centering
    {\begin{tabular}{cc|c}
            \hline
            & Parameter & Value \\
            \hline\hline
                            & $f_{K^*}$ & $220 \pm 5$ MeV \\
            Decay constants \cite{Cerqueira:2015vva,Rodrigues:2017qsm}
                            & $f_{D}$   & $206.7\pm 8.5 \pm 2.5$ MeV \\
                            & $f_{D_s}$ & $257.5 \pm 6.1 $ MeV\\\hline
                            & $m_{K^*}$ & $0.892$ GeV \\
            Meson masses \cite{Agashe:2014kda}
                            & $m_{D}$   & $1.869$ GeV \\
                            & $m_{D_s}$     & $1.968$ GeV \\\hline
                            & $m_c$     & $1.27\pm 0.025$ GeV \\
            Quark masses \cite{Agashe:2014kda,Nakamura:2010zzi}
                            & $m_{u,d}$ & $0$ GeV\\
                            & $m_s$     & $101 +29 - 21$ MeV \\\hline
            & $\left < q\bar{q} \right>$           & $-(230\pm 30)^3 \,\rm{MeV}^3$\\
            Condensates \cite{Rodrigues:2017qsm}
                            & $\left < s\bar{s} \right>$           & $-(290\pm 15)^3\, \rm{MeV}^3$\\
            & $\left < g^2G^2 \right >$            &$0.88\pm 0.16\, \rm{GeV}^4$\\
            &$\left<q g\sigma G \bar{q} \right >$ & $(0.8\pm 0.2) \left < q\bar{q} \right>\,\rm{GeV}^5$\\ \hline
\end{tabular}}{\caption{\label{numvalues} Physical parameters.}}
\end{table}

\begin{figure}[h!]
    \subfloat[ $D$ meson, structure $p_\mu$]{\includegraphics[width = 0.47\textwidth]{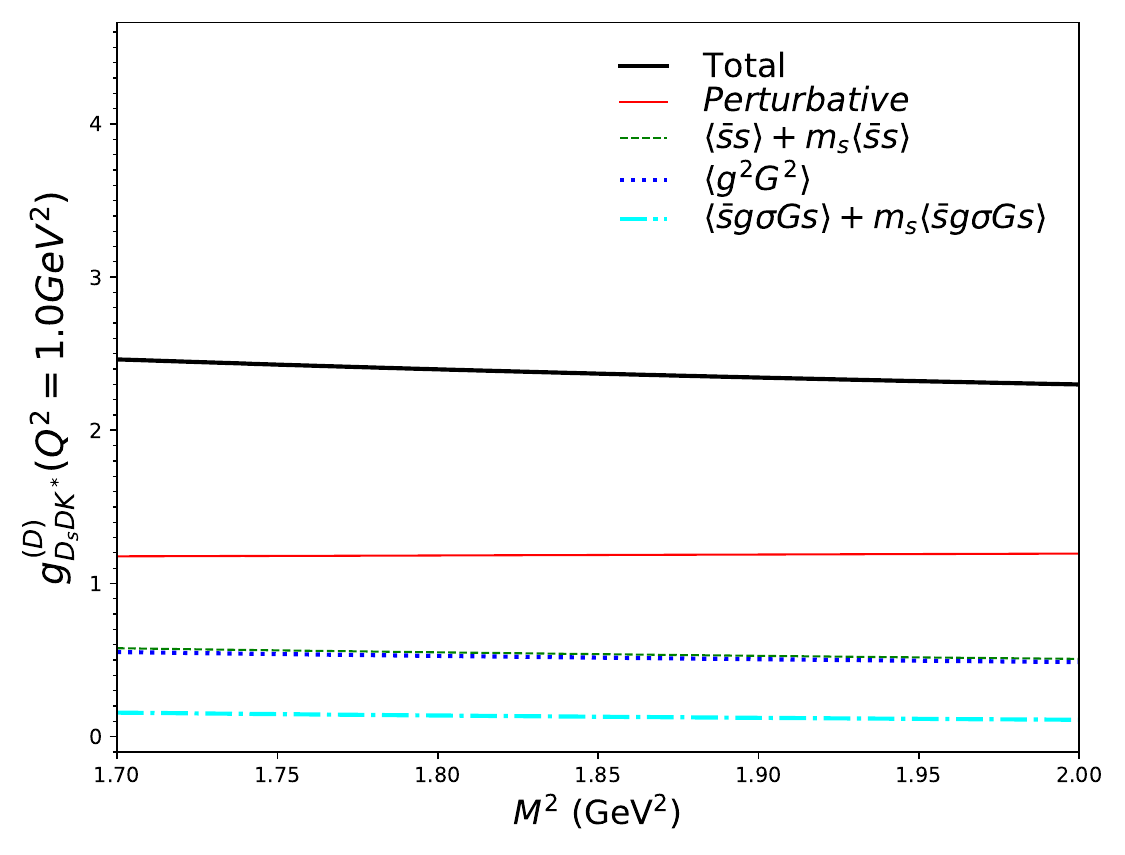}}\qquad
    \subfloat[$D_s$ meson, structure $p_\mu$]{\includegraphics[width = 0.47\textwidth]{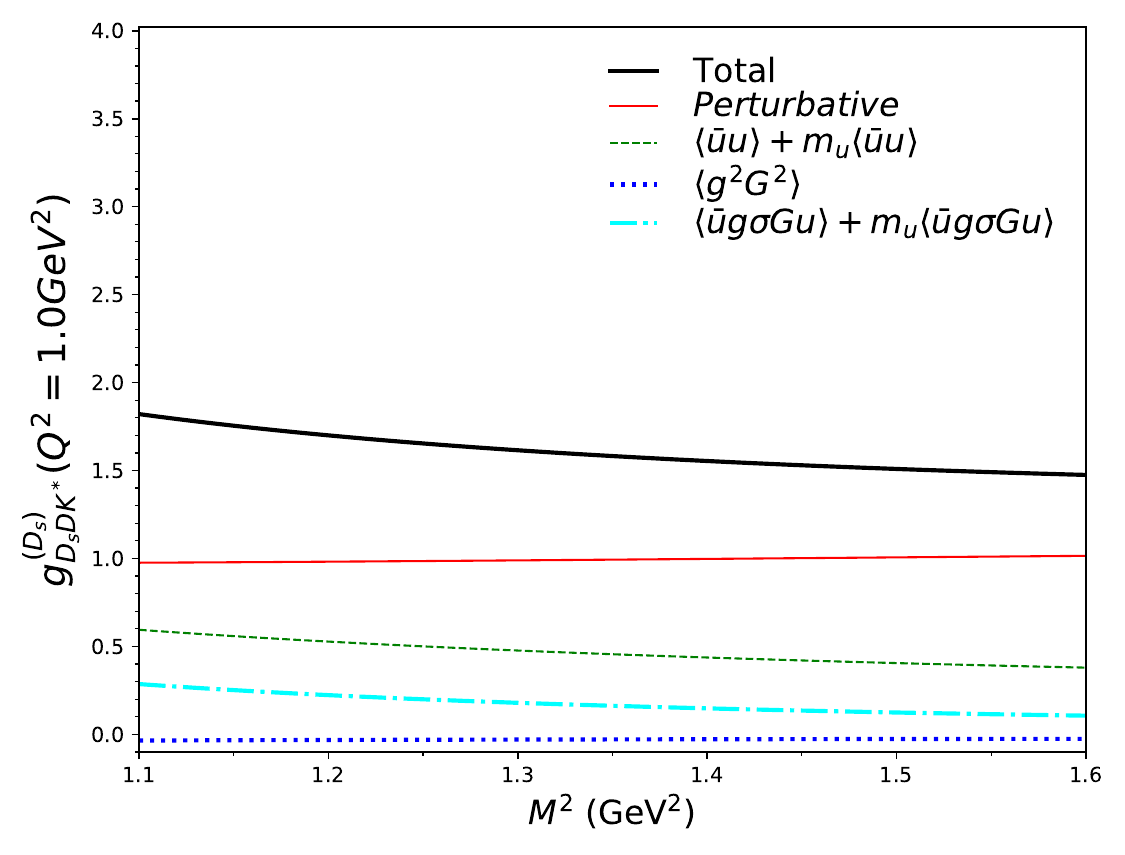}}\qquad
    \subfloat[$K^*$ meson, structure $p_\mu^\prime$]{\includegraphics[width = 0.47\textwidth]{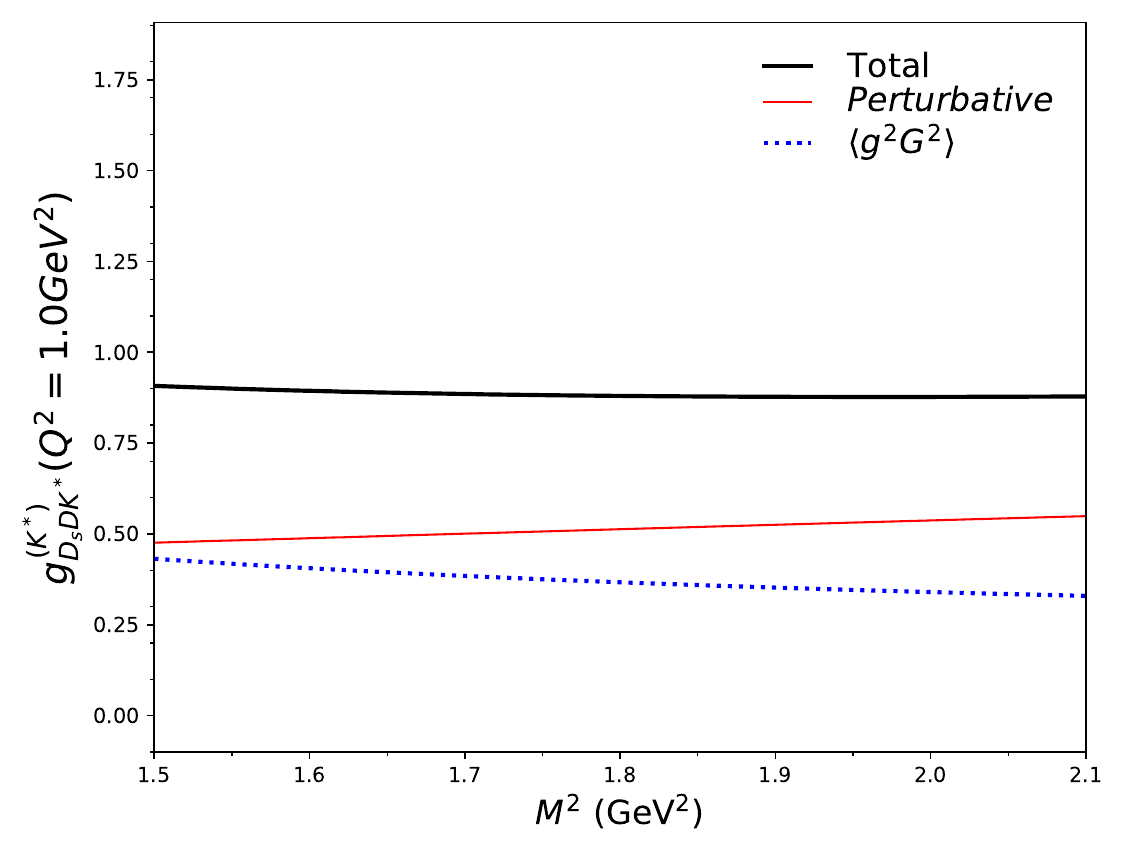}}
    \caption{OPE stability and convergence,  for each off-shell case and structures, with $Q^2=1\,$GeV$^2$.}
    \label{opefig}
\end{figure}

\begin{figure}[h!]
    \subfloat[$D$ off-shell, structure $p_\mu$]{\includegraphics[width = 0.47\textwidth]{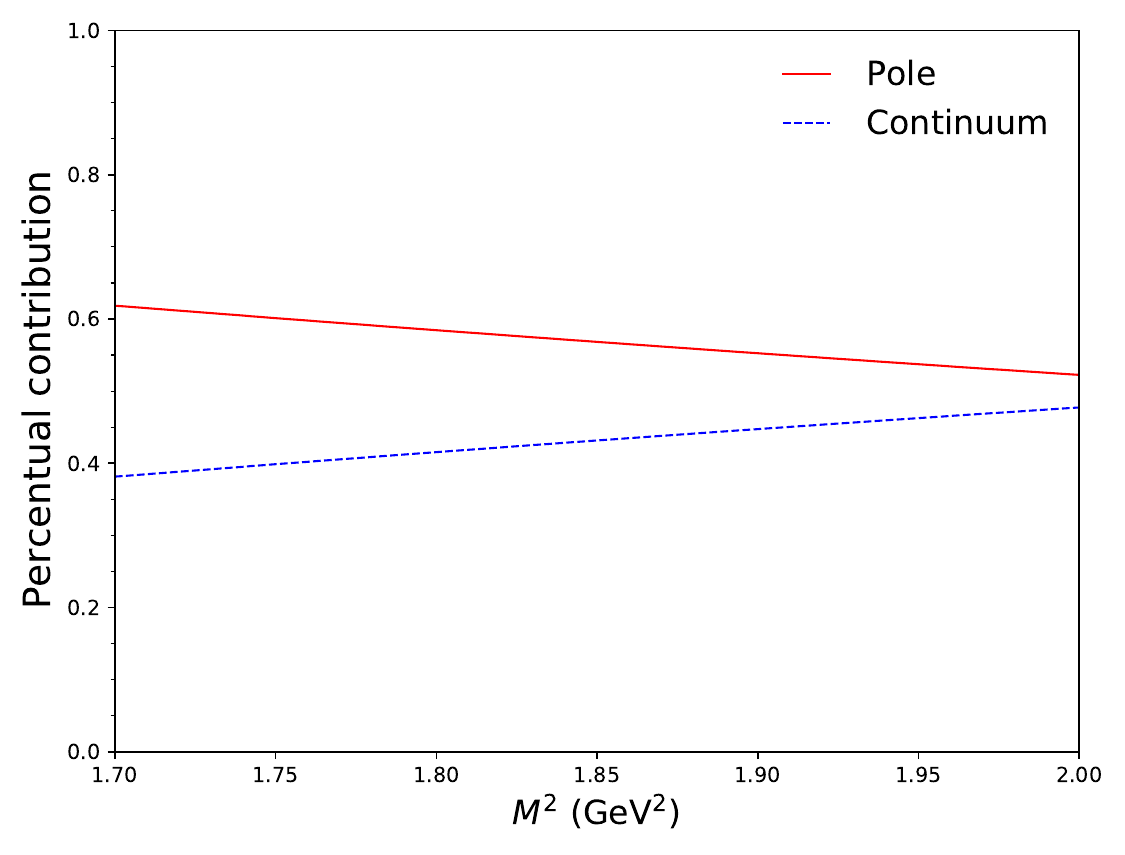}}\qquad
    \subfloat[$D_s$ off-shell, structure $p_\mu$]{\includegraphics[width = 0.47\textwidth]{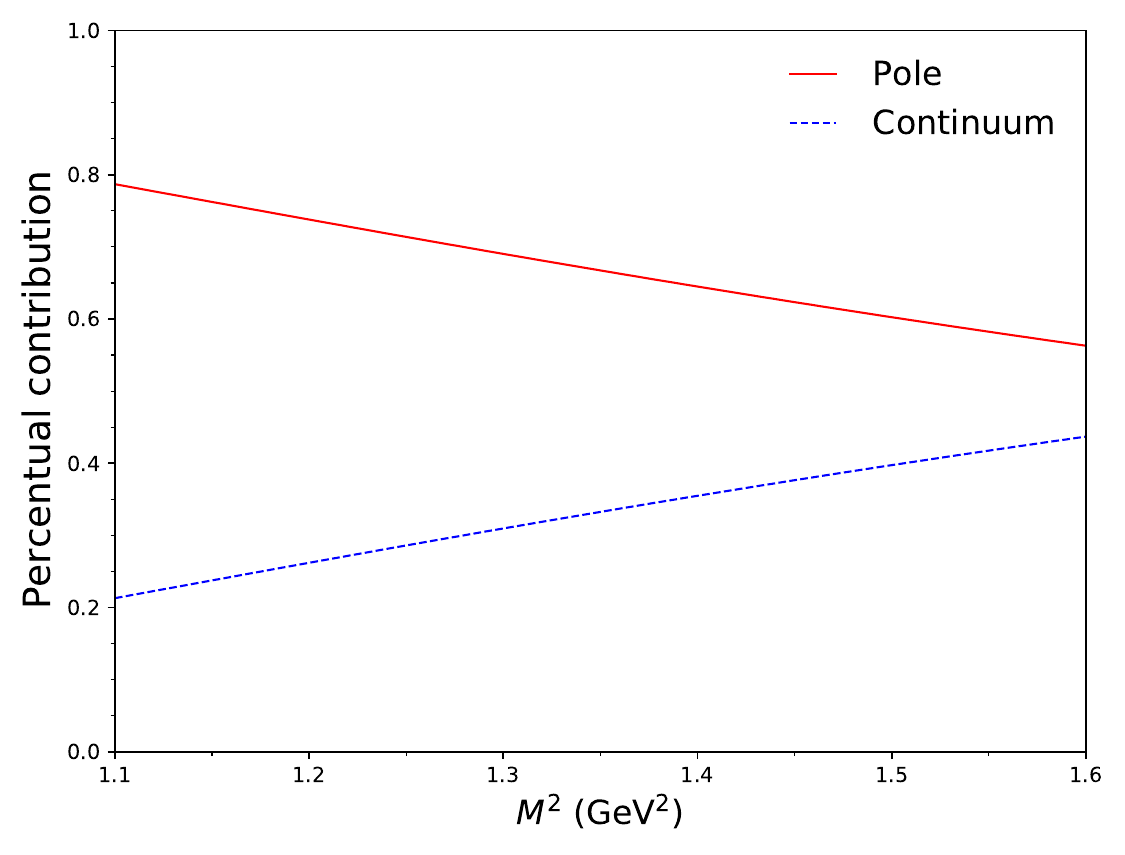}}\qquad
    \subfloat[$K^*$ off-shell, structure $p_\mu^\prime$]{\includegraphics[width = 0.47\textwidth]{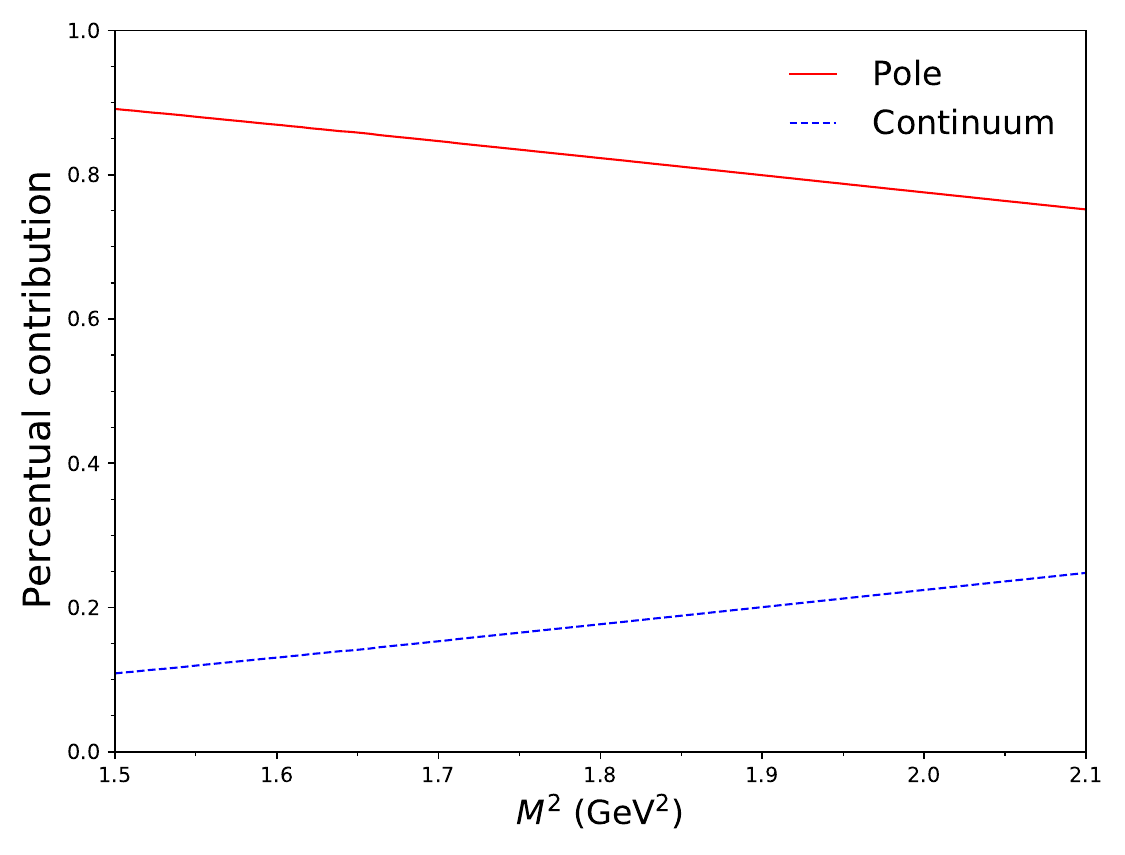}}\\
    \caption{Pole-continuum contributions, for each off-shell case and structures, with $Q^2=1\,$GeV$^2$.}
    \label{polecontinuum}
\end{figure}

The values of the hadronic parameters used in the numerical
calculations of the sum rules are listed on Table \ref{numvalues}.
We work with the  $p_\mu$  structure for the $D$ and $D_s$
off-shell, and with the $p_\mu^\prime$ structure for the $K^*$
off-shell. This choice is based on which one of the structures
provides the best fit for the form factors while respecting the
constraints imposed by the QCDSR. We first show the OPE
convergence in Fig.~(\ref{opefig}), where it is possible to
observe that each term of the OPE series is smaller than the
previous ones. It is also possible to check the region of
stability of the total OPE series in this figure. The
pole-continuum comparison is shown next in
Fig.~(\ref{polecontinuum}), establishing the region with pole
dominance over the continuum contributions. From the
Figs.~(\ref{opefig}) and (\ref{polecontinuum}), we extract
the Borel mass windows, i.e., the range of values of the
Borel mass in which the sum rules are reliable.  The Borel
window is further decreased with the fit of the form factors, that
we will present next.

\begin{figure}[h!]
    \centering
    {\includegraphics[width = 0.62\textwidth]{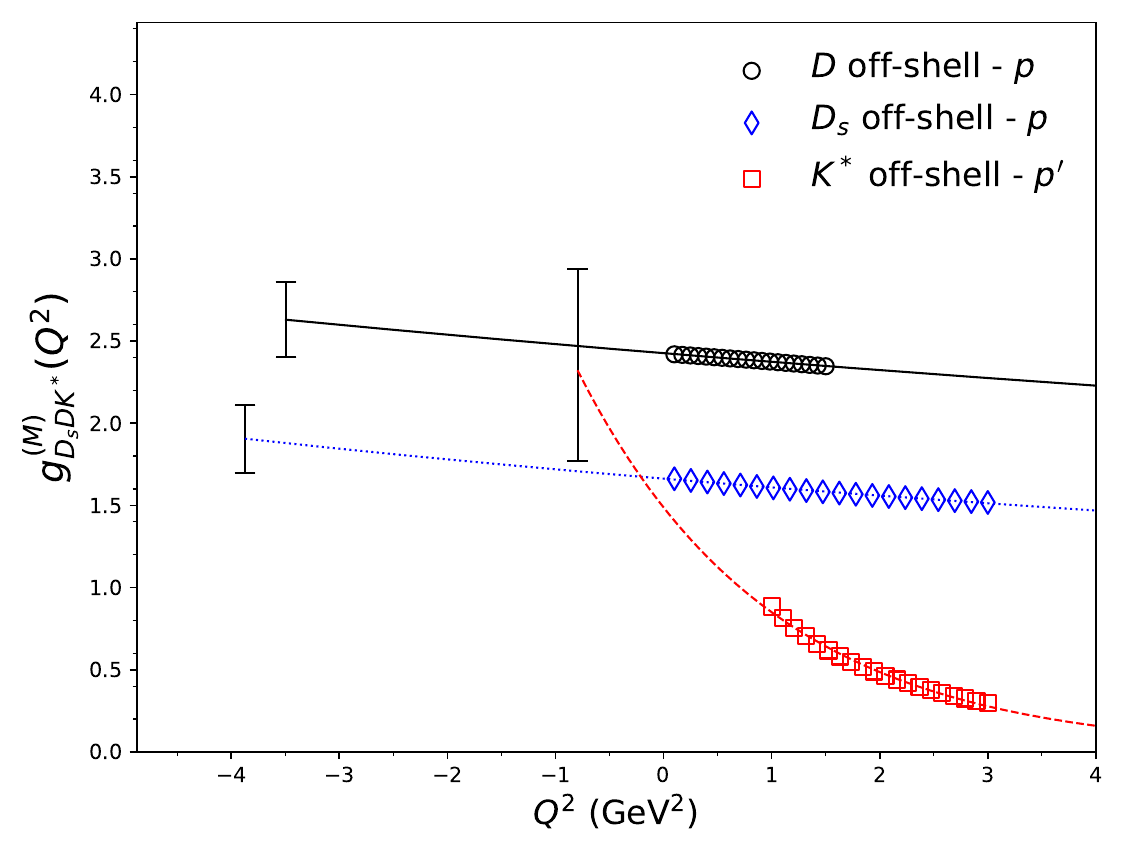}}
    \caption{Form factors of the vertex $D_sDK^*$, and their respective extrapolated functions. The values of the form factor at the off-shell meson pole and the error bars are also indicated.}
    \label{formfactor}
\end{figure}

The Fig.~(\ref{formfactor}) shows the resultant 
numerical calculations of the form factors from Eq.~(\ref{formfactorfunction}). The
form factors were fitted to the numerical data
computed as either a monopolar function,
\begin{equation}
g(Q^2)=\frac{\mathcal{A}}{\mathcal{B}+Q^2}\,,
\end{equation}
or an exponential function,
\begin{equation}
g(Q^2)=\mathcal{A}\,e^{-\mathcal{B}Q^2}\,,
\end{equation}
where the parameters $\mathcal{A,B}$ were determined by the fitting procedure. The form factors are chosen according to the best fit of each form factor to the numerical data. The results  are shown in Table~\ref{gparameters}.

The $\Delta$ parameters, the Borel mass $M^2$, $Q^2$ windows and also the coupling constants results for the $M$  are shown in Table~\ref{tab1} for each off-shell meson case. The values are obtained by the extrapolation of the fitted curves of Table~\ref{gparameters} to the non-Euclidean region $Q^2<0$, and then calculating the value of the form factor at the off-shell meson pole. The result thus obtained for the vertex $D_sDK^*$ coupling constant is taken from
\begin{equation}
g_{D_sDK^*}^{(M)}=g_{D_sDK^*}^{(M)}(Q^2=-m_M^2)\,.
\end{equation}

The coupling constant error bars are presented in
Fig.~(\ref{formfactor}). By comparing the error bars of the cases,
it is possible to see that their values are compatible. The final
result for the coupling constant of the vertex is taken from the
mean value of the coupling constants presented in
Table~\ref{tab1}:
\begin{equation}
g_{D_sDK^*} = 2.29^{+0.65}_{-0.41}.\label{gfinal}
\end{equation}

\begin{table}[h!]
	\centering
	{\begin{tabular}{p{4cm}p{3cm}p{3cm}p{2cm}p{2cm}}\hline
			Form factor        & Structure   & Type  & $\mathcal{A}$ &$\mathcal{B}$ \\\hline
			$g^{(D)}_{DsDK^*}(Q^2)$    &  $p_\mu$     & Monopolar      & 109.52 & 45.140\\
			$g^{(K^*)}_{DsDK^*}(Q^2)$  &  $p^\prime_\mu$ & Exponential & 1.488  & 0.560 \\
			$g^{(D_s)}_{DsDK^*}(Q^2)$  &  $p_\mu$  & Monopolar         & 50.431 & 30.332 \\ \hline
	\end{tabular}}\caption{\label{gparameters} Numerical results for the parametrization of the form factors.}
\end{table}

\begin{table}[h!]
    \centering
        \begin{tabular}{p{3cm}>{\centering}p{3cm}>{\centering}p{3cm}>{\centering\arraybackslash}p{3cm}}
            \cline{2-4}
            \multirow{2}{*}{}&\multicolumn{3}{c}{Off-shell meson / structure}\\\cline{2-4}
            &  {$D$ / $p_\mu$}   &  {$D_s$ / $p_\mu$}  &    {$K^*$} / $p^\prime_\mu$\\
            \hline\hline
            {$\Delta_s$ (GeV) }        & 0.5 & 0.5 &   0.6\\
            {$\Delta_u$ (GeV) }       & 0.6 & 0.6 &  0.6 \\
            {$M^2$ (GeV$^2$) }             & $\left [ 1.7, 2.0 \right ]$ & $\left [ 1.1 , 1.6 \right ]$ &   $\left [ 1.5, 2.1 \right ]$  \\
            {$Q^2$ (GeV$^2$) }           & $\left [ 0.1, 1.5 \right ]$ & $\left [ 0.1, 3.0 \right ]$ &   $\left [ 1.0, 3.0 \right ]$   \\
            \hline
            \hline
            {$g^{(M)}_{D_sDK^*}$} & $2.63^{ + 0.23 }_{ - 0.23 }$ & $1.91^{ + 0.20 }_{ - 0.21 }$ &    $2.32^{+0.62}_{-0.55}$ \\
            \hline
    \end{tabular} \caption{\label{tab1} Numerical results for the  $\Delta$ parameters, the $M^2$  and $Q^2$ windows and coupling constant for each off-shell case.  }
\end{table}

\newpage

\newpage

\subsection{Error analysis}

The percentage deviations for each parameter of the QCDSR
calculation are presented in Table~\ref{error}.  Such deviations are obtained by varying each parameter presented in the QCDSR calculation within their respective uncertainty windows, obtaining their respective coupling constants and then comparing the value of the coupling constant for each of the extremes of these windows to the values of the coupling constants presented in Table III.

In the case of decay constants, quark masses and condensates, the percentage deviations are calculated from the variation of such parameters within their respective uncertainties presented in Table I. For the parameter $Q^2$, we obtain the deviation by making variations of 20\% in the window widths shown in Table III. For the threshold parameters ($\Delta_s$ and $\Delta_u$), we studied variations of $\pm 0.1GeV$ in each one, both separately and together. For the Borel mass $M^2$, we calculated the standard deviation of the coupling constant within the Borel windows shown in Table III. Finally, the deviations in the parameters of the fit functions ($\mathcal{A}$ and $\mathcal{B}$) are obtained by varying those parameters  within their uncertainties indicated by the fit method itself.

The percentage deviations therefore indicate the impact that the variation of each parameter has on the calculation of the vertex coupling constant. From the Table~\ref{error}, it is possible to check that the main sources of errors are the gluon condensate and the strange and charm quarks' masses for the $K^*$ off-shell. For the $D$ and $D_s$ off-shell cases, the main sources of errors are the $\Delta$'s, the decay constant of meson $D$ and the quark condensates .

The chosen
structures ($p_\mu$ for $M=D,D_s$, $p^\prime_\mu$ for $M=K^*$) are
determined after a rigorous error analysis of the numerical fit of
the form factors. The constrains on the numerical fit were the
following:
\begin{itemize}
    \item Minor global error;
    \item Less sensitivity to the fit, quarks masses and condensates;
    \item Similarity of the form factor behavior between the $D$ and $D_s$ off-shell cases;
    \item Larger as possible $Q^2$ window;
    \item Stable Borel windows;
    \item A convergent OPE series, with the perturbative term contribution bigger than the sum of the condensates contributions;
    \item Pole contribution must be bigger than the continuum contribution.
\end{itemize}

\begin{table}[h!]
    \centering{
    \begin{tabular*}{\linewidth}{c|>{\centering}p{3.5cm}>{\centering}p{2.7cm}>{\centering\arraybackslash}p{2.7cm}}
        \hline
        {\textbf{Parameter}}  & $\Delta g_{D_s D K^*}^{(D - p)}$(\%) & $\Delta g_{D_s D K^*}^{(D_s - p)}$(\%) &  $\Delta g_{D_s D K^*}^{(K^* - p')}$(\%) \\
        \hline\hline
                $M^2$ & 1.18 & 0.55 &  7.67\\
        $Q^2$ & 0.18 & 0.60 &  13.92\\
        $\Delta_s \pm 0.1 (GeV), \Delta_u \pm 0.1 (GeV)$ & 5.44 & 5.83 &  5.78\\
        $\left<\bar{q}g\sigma\cdot G q \right>$ & 0.19 & 0.03 &  - \\
        $\left<\bar{q}q\right>$ & 2.30 & 6.73 &   - \\
        $\left<g^2G^2\right>$ & 2.75 & 0.62 & 7.09\\
        $\text{Fitting Parameters } (\mathcal{A} \text{ and } \mathcal{B})$ & 1.42 & 2.51 &  0.82\\
        $f_{D_s}$ & 1.94 & 1.94 & 1.94\\
        $f_{D}$ & 4.35 & 4.35 &  4.36\\
        $f_{K^*}$ & 1.85 & 1.85 &  1.85\\
        $m_c$ & 2.02 & 0.93 &  2.31\\
        $m_s$ & 0.83 & 1.17 &  15.44\\\hline
    \end{tabular*} {\caption{Percentage deviation of the coupling constants according with the variation of
    each parameter for the structures $p_\mu$ ($D$ and $D_s$ off-shell cases) and $p'_\mu$ ($K^*$ off-shell). \label{error}}}}
\end{table}

\section{Summary and conclusions}

In this work, the QCDSR technique was applied to perform the
calculation of the form factors and coupling constant for the
charmed meson vertex $D_sDK^*$. The form factors for each one of
the off-shell mesons were computed, so that the
uncertainties were minimized. The parametrization used for the
form factors were either monopolar or exponential functions, as
usual for this type of vertex. The values of the form factors at
the respective off-shell meson pole mass ($Q^2\to -m_{\rm off}^2$)
were combined to obtain the coupling constant of the vertex, and
the final result was presented in Eq.~(\ref{gfinal}).

\begin{table}[h!]
    \centering
    {\begin{tabular}{p{4cm}>{\centering\arraybackslash}p{2.7cm}}\hline
            Reference & $g_{D_sDK^*}$ \\\hline
            This Work   &  $2.29^{+0.65}_{-0.41}$ \\
			SU(4)  \cite{Azevedo:2003qh}     & 5\\
            LCSR   \cite{Wang:2007zm}      &   $1.61\pm0.62$   \\
            QCDSR \cite{Janbazi:2017mpb}  & $3.26\pm0.43$  \\\hline
    \end{tabular}}\caption{\label{tab4} Comparison of the coupling constant of the vertex  $g_{D_sDK^*}$ that were obtained in different works.}
\end{table}

We can compare our results with previous calculations for the same
vertex, as shown in Table~\ref{tab4}. In Ref.~\cite{Azevedo:2003qh}, the $SU(4)$ estimate for the coupling constant is more than two times bigger than ours, showing a large $SU(4)$ breaking effect. In Ref.~\cite{Wang:2007zm}, the coupling constant of the
vertex $D_sDK^*$ was computed in the Light-cone Sum Rules (LCSR)
approach. Their result presents a mean value $30$\% smaller than
ours, but compatible within the error bars. There is also
a previous QCDSR calculation in Ref.~\cite{Janbazi:2017mpb} in which the result of the
coupling constant  is also compatible with ours, within the error
bars. Our mean value for the coupling constant is 30\%
smaller than the result presented in Ref.~\cite{Janbazi:2017mpb}.  Despite using QCDSR method, as in our work, there are somewhat important differences between the calculations that can account for such divergence between our result and Ref.~\cite{Janbazi:2017mpb} result:

\begin{itemize}
	\item while we calculated all three off-shell cases, three form factors, only the  $D$ and $K^*$ off-shell cases were calculated in Ref.~\cite{Janbazi:2017mpb}. As can be seen from Fig.~(\ref{formfactor}), the $D_s$ off-shell case presents a smaller coupling constant than the other two cases, making the final result of the coupling constant smaller than if only the $D$ and $K^*$ off-shell cases were considered;
		
	\item we include the gluon condensates in the $D$ off-shell case. In Fig.~(\ref{opefig}), it is shown that terms with gluon condensates are equally important to the OPE as terms with quark condensates for the $D$ off-shell calculation and, therefore, can not be neglected; 
	
	\item we analyze the pole-continuum dominance and for these reason our window for $Q^2$ momentum is very different of Ref.~\cite{Janbazi:2017mpb}. This guarantees that the continuum is not bigger that the pole contribution, where also the OPE series could not be convergent;  
	
	\item  the numerical values of mass parameters are different than ours, and there is no information for the values of the condensates. As shown in Table~\ref{error}, the $K^*$ off-shell case is particularly sensible to the strange quark mass and the gluon condensate;
	
	\item finally, in Ref.~\cite{Janbazi:2017mpb}, the form factor fit was made using a Gaussian function for both off-shell cases: for the $K^*$ off-shell, the data do not decrease so fast and do not fit as well as for the $D$ off-shell case. If it were used a monopolar parametrization instead, the extrapolation of the form factors would give a higher value for the coupling constant.
	
\end{itemize}


\appendix
\section{Perturbative and condensate contributions to the OPE}

We list the expressions for the OPE terms. For concision, the huge gluon condensates expressions are omitted in this paper.

\subsection{$D$ off-shell}

The vertex for the $D$ off-shell case, shown in Fig.~(\ref{diagrams}a) is obtained using the proper currents of Eq.~(\ref{3currents}) on the correlator of Eq.~(\ref{correlator}): $j_1=j_{D_S}$, $j_2=j_{D}$ and $j_3=j_{K^*}$. On the phenomenological side, the quadrimomenta in Eqs.~(\ref{massdecay})  are: $p_a=q$ for the $D$; $p_b=p^\prime$ for the $D_s$ and $p_c=p$ for the $K^*$. Also, the amplitude matrix taken from the lagrangian of Eq.~(\ref{lagrangian}) is the first one in the Eq.~(\ref{matrices}).

The result for the phenomenological expression is:

\begin{equation}
\Gamma^{phen}_\mu=\frac{g_{D_s DK^*}^{(D)}(Q^2)f_D f_{D_s} f_{K^*} \frac{m_D^2
        m_{D_s}^2}{m_{K^*}m_c(m_c+m_s)}[(m_{K^*}^2+m_{D_s}^2-q^2)p_{\mu}-2m_{K^*}^2
    p'_{\mu}]}{(p-m_{K*}^2)(p'^2-m_{D_s}^2)(q^2-m_D^2)}\,.
\end{equation}

On the OPE side, for the $p_\mu$ structure, besides the perturbative term, there are contributions from the quark, gluon and mixed condensates. For the $p^{\prime}_\mu$ structure, there is only the perturbative contribution. The expression for the each contribution are listed bellow.

\begin{itemize}
    \item Perturbative term:
    \begin{eqnarray}
    DD \left[ \Gamma^{pert}_\mu \right]=\frac{3}{\sqrt{\lambda}}\left\{ [A \cdot C-m_s m_c -p' \cdot k+m_s^2] p_{\mu} + [B\cdot C+p \cdot
    k-m_s^2]p'_{\mu}\right\} \,,
    \end{eqnarray}
    where:
    \begin{equation}
    A=2\pi\left[\frac{\bar{k}_0}{\sqrt{s}}-\frac{p_0^\prime\overline{\vert\vec{k}\vert}}{\sqrt{s}|\vec{p}|}{\overline{\cos\theta}}\right]\,,\qquad B=2\pi\frac{\overline{\vert\vec{k}\vert}}{|\vec{p}|}{\overline{\cos\theta}}\,,\label{AB}
    \end{equation}
    \begin{equation}
    C = (-p' \cdot p +m_s m_c+2p' \cdot k-m_s^2)\,,
    \end{equation}
    \begin{equation}
    s=p^2;\quad u=p^{\prime 2};\quad t=q^2\,,
    \end{equation}
    \begin{equation}
    k^2=m_s^2\,,
    \end{equation}
    \begin{equation}
    k_0=\frac{s+m_s^2}{2\sqrt{s}}\,,
    \end{equation}
    \begin{equation}
    \overline{|{\vec k}|}^2=k_0^2-m_s^2\,,
    \end{equation}
    \begin{equation}
    \overline{\cos \theta}=\frac{2p'_0k_0+m_c^2-m_s^2-u}{2 |\vec{p'}|\,
        |\vec{k}|}
    \end{equation}
    \begin{equation}s > m_s^2\,,
    \end{equation}
    \begin{equation}
    u > t+m_s^2\,.
    \end{equation}

    \item Quark condensate $\langle\bar{s}s\rangle$:
    \begin{equation}
    \mathcal{B}\mathcal{B}[\Gamma^{\langle s \bar s \rangle}_\mu]= -\langle s \bar s \rangle [m_c p_{\mu}]e^{-m_c^2/M'^2}\,.
    \end{equation}

    \item Quark condensate $m_s\langle\bar{s}s\rangle$:
    \begin{multline}
    \mathcal{B}\mathcal{B} \left [ \Gamma^{m_s\left<\bar{s}s\right>}_\mu \right ] = -\frac{m_s \left < s\bar{s} \right >}{2M^2M'^2} p_\mu e^{-\frac{m_c^2}{M'^2}}
   \left[ (m_c^2+M'^2)M^2  - (M^2-Q^2-m_c^2)M'^2\right]\,.
    \end{multline}

    \item Mixed condensate $\langle\bar{s}g\sigma Gs\rangle$:
    \begin{multline}
    \mathcal{B}\mathcal{B} \left [ \Gamma^{\left< \bar{s}g\sigma Gs\right>}_\mu \right ] = -\frac{2m_c\left< \bar{s}g\sigma Gs\right>}{8M^4M'^4} p_\mu  e^{-\frac{m_c^2}{M'^2}}\left[ m_c^2M^4  + M^4M'^2 \right.\\ \left. + (Q^2 + m_c^2)M^2M'^2 - M^2M'^4     \right ]\,.
    \end{multline}

    \item{Mixed condensate $m_s\langle\bar{s}g\sigma Gs\rangle$}:
    \begin{multline}
    \mathcal{B}\mathcal{B} \left [ \Gamma^{m_s \left< \bar{s}g\sigma G s\right>}_\mu \right ] = \frac{m_s\left< \bar{s}g\sigma Gs\right>}{24M^6M'^6} p_\mu  e^{-\frac{m_c^2}{M'^2}}\left[ 6M^{10}+2m_c^4M^6-34M^6M'^4 \right.\\
    -15m_c^2M^6M'^2 + 6m_c^2(m_c^2+Q^2)M^4M'^2 + 2(Q^2-6m_c^2)M^4M'^4\\
    \left.+3(Q^2+3m_c^2)M^2M'^6 + 6M^4M'^6     \right ]\,.
    \end{multline}
\end{itemize}


\subsubsection{$D_s$ off-shell}

The vertex for the $D_s$ off-shell case, shown in Fig.~(\ref{diagrams}b) is obtained using the proper currents of Eq.~(\ref{3currents}) on the correlator of Eq.~(\ref{correlator}): $j_1=j_{D}$, $j_2=j_{D_s}$ and $j_3=j_{K^*}$. On the phenomenological side, the quadrimomenta on relations of Eq.~(\ref{massdecay}) are: $p_a=p^\prime$ for the $D$; $p_b=q$ for the $D_s$ and $p_c=p$ for the $K^*$. The amplitude matrix taken from the lagrangian of Eq.~(\ref{lagrangian}) is the second one in Eq.~(\ref{matrices}).

The result for the phenomenological expression is:

\begin{equation}
\Gamma^{phen}_\mu=\frac{-g_{D_s DK^*}^{(D_s)}(Q^2)f_D f_{D_s} f_{K^*} \frac{m_D^2
        m_{D_s}^2}{m_{K^*}m_c(m_c+m_s)}[(m_{K^*}^2+m_D^2-q^2)p_{\mu}-2m_{K^*}^2
    p'_{\mu}]}{(p-m_{K*}^2)(p'^2-m_D^2)(q^2-m_{D_s}^2)}
\end{equation}

On the OPE side, there are contributions for the $p_\mu$ and $p_\mu^\prime$ structures from perturbative, quark and gluons condensates. For the $p_\mu$ structure there is also the contribution from the mixed condensate. The expressions for the each OPE contribution are listed bellow.

\begin{itemize}

    \item Perturbative:

    \begin{equation}
    DD \left[\Gamma^{pert}_\mu \right]=\frac{3}{\sqrt{\lambda}}\left\{[A\cdot C +p' \cdot k] p_{\mu} +[B\cdot C-p \cdot k]p'_{\mu}\right\}\,,
    \end{equation}
    where the functions $A$ and $B$ are the same as in Eq.(\ref{AB}), and other parameters are given bellow:
    \begin{equation}
    C = (p' \cdot p -m_sm_c-2p'\cdot k)\,,
    \end{equation}
    \begin{equation}
    k^2=0\,,
    \end{equation}
    \begin{equation}
    k_0=\frac{s-m_s^2}{2\sqrt{s}}\,,
    \end{equation}
    \begin{equation}
    \overline{|{\vec k}|}^2=k_0^2\,,
    \end{equation}
    \begin{equation}
    \overline{\cos \theta}=\frac{2p'_0k_0+m_c^2-u}{2 |\vec{p'}| |\vec{k}|}\,,
    \end{equation}
    \begin{equation}
    s > m_s^2\,,
    \end{equation}
    \begin{equation}
    u > t-m_s^2\,.
    \end{equation}

    \item Quark condensates $ \langle q \bar q \rangle$:
    \begin{equation}
    \mathcal{B}\mathcal{B}[\Gamma_\mu^{\langle\bar q  q \rangle}]= \langle \bar q  q \rangle [m_c
    p_{\mu} - m_s p'_{\mu}]e^{-m_s^2/M^2}e^{-m_c^2/M'^2}\,.
    \end{equation}
    \item Mixed condensate $\langle\bar{q}g\sigma Gq\rangle$:
    \begin{multline}
    \mathcal{B}\mathcal{B} \left [ \Gamma^{\left< \bar{q}g\sigma Gq\right>}_\mu \right ] = \frac{\left< \bar{q}g\sigma Gq\right>}{8M^4M'^4} p_\mu e^{-\frac{m_s^2}{M^2}} e^{-\frac{m_c^2}{M'^2}}\left[ 2m_c^3M^4 + 2m_cm_sM'^4 \right.\\ \left. + 2m_cM^4M'^2 + 2m_c(Q^2 + m_s^2 + m_c^2)M^2M'^2 - 2(m_s+m_c)M^2M'^4     \right ]\,.
    \end{multline}
\end{itemize}

\subsubsection{$K^*$ off-shell}

The vertex for the $K^*$ off-shell case, shown in Fig.~(\ref{diagrams}c), is obtained using the proper currents of Eq.~(\ref{3currents}) on the correlator of Eq.~(\ref{correlator}): $j_1=j_{D}$, $j_2=j_{K^*}$ and $j_3=j_{D_s}$. On the phenomenological side, the quadrimomenta on the Eq.~(\ref{massdecay}) relations are: $p_a=p^\prime$ for the $D$; $p_b=p$ for the $D_s$ and $p_c=q$ for the $K^*$. The amplitude matrix taken from the lagrangian of Eq.~(\ref{lagrangian}) is the third one in Eq.~(\ref{matrices}).

The phenomenological side is given by:
\begin{multline}
\Gamma_\mu^{phen}=- \frac{g_{D_s DK^*}^{(K^*)}(Q^2)f_D f_{D_s} f_{K^*}\frac{ m_D^2
        m_{D_s}^2}{m_{K^*}m_c(m_c+m_s)}}{(p-m_{D_s}^2)(p'^2-m_D^2)(q^2-m_{K^*}^2)}\times [(m_{D_s}^2-m_{K^*}^2-m_D^2)p_{\mu}\\
        -(m_D^2-m_{D_s}^2-m_{K^*}^2)
            p'_{\mu}]\,.
\end{multline}

On the OPE side, there are contributions for the $p_\mu$ and $p_\mu^\prime$ structures from perturbative and gluon condensates. The expressions for the  OPE terms are listed bellow.

\begin{itemize}
    \item  Perturbative:
    \begin{equation}
    DD\left[\Gamma^{pert}_\mu \right]=\frac{3}{\sqrt{\lambda}}\left\{[A\cdot C+m_c^2-p' \cdot k] p_{\mu} +[B\cdot C+m_c^2-m_c m_s-p \cdot k]p'_{\mu}\right\}\,,
    \end{equation}
    where the functions $A$ and $B$ are the same as in Eq.(\ref{AB}), and other parameters are listed bellow:
    \begin{equation}
    C = (p' \cdot p +m_c m_s- m_c^2)\,,
    \end{equation}
    \begin{equation}
    k^2=m_c^2\,,
    \end{equation}
    \begin{equation}
    k_0=\frac{s+m_c^2-m_s^2}{2\sqrt{s}},,
    \end{equation}
    \begin{equation}
    |\bar{\vec k}|^2=k_0^2-m_c^2,,
    \end{equation}
    \begin{equation}
    \overline{\cos \theta}=\frac{2p'_0k_0-m_c^2-u}{2 |\vec{p'}| |\vec{k}|},,
    \end{equation}
    \begin{equation}
    s > m_c^2-m_s^2,,
    \end{equation}
    \begin{equation}
    u > t+m_c^2-m_s^2,.
    \end{equation}

\end{itemize}

\renewcommand{\bibsection}{\subsection*{REFERENCES}}
\bibliography{bibliografia}

\end{document}